\documentclass[twocolumn,preprint]{aastex62}
\usepackage[utf8x]{inputenc}
\usepackage{amsmath,amsfonts,amssymb,mathrsfs,bm,enumerate,graphicx}
\usepackage[T1]{fontenc}
\usepackage[all]{hypcap}
\hypersetup{
    colorlinks=true,
    linkcolor=magenta,
    filecolor=magenta,      
    urlcolor=blue,
}

\usepackage{ulem}
\usepackage{threeparttable}
\usepackage{ae,aecompl}

\def\red#1 {\textcolor{red}{#1}\ }   
\def\blue#1 {\textcolor{blue}{#1}\ }   
\def\purple#1 {\textcolor{purple}{#1}\ }   

\newcommand{\ba}{\begin{eqnarray}}
\newcommand{\ea}{\end{eqnarray}}

\shorttitle{Kozai Migration Around White Dwarfs}
\shortauthors{Mu\~noz \& Petrovich}

\begin{document}

\title{\bf Kozai Migration Naturally Explains the White Dwarf Planet WD1856b}
\author[0000-0003-2186-234X]{Diego J. Mu\~noz}
\affiliation{CIERA, Northwestern University, 1800 Sherman Ave.,Evanston, IL 60208, USA}
\author[0000-0003-0412-9314]{Cristobal Petrovich}
\affiliation{Steward Observatory, University of Arizona, 933 N. Cherry Ave., Tucson, AZ 85721, USA}
\affiliation{Instituto de Astrofísica, Facultad de Física, Pontificia Universidad Católica de Chile, Casilla 306, Santiago 22, Chile}

\begin{abstract}
The Jovian-sized object WD~1856~b transits a white dwarf (WD) in a compact $1.4$-day orbit. Unlikely to have endured stellar evolution in
its current orbit, WD~1856~b 
is thought to have migrated from much wider separations.
Because the WD is old, and a member of a well-characterized hierarchical multiple, the well-known 
Kozai mechanism provides an effective migration channel for
WD~1856~b.
Moreover, the lack of tides in the star allows us to directly connect
the current semi-major axis to the pre-migration one, from which we can 
infer the initial conditions of the system.
By further demanding that successful migrators survive all previous phases of stellar evolution, we are able to constrain the mass of WD~1856~b to be $\simeq0.7-3M_{\rm J}$ and  its main sequence semi-major axis to be $\simeq 2-2.5$ au. These properties
imply that WD~1856~b was born a typical gas giant.
We further estimate the occurrence rate of  Kozai-migrated planets around WDs to be ${\cal O}(10^{-3}{-}10^{-4})$, suggesting that 
WD~1856~b is the only one in the {\it TESS} sample, but implying
${\cal O}(10^2)$ future detections by LSST. In a sense,  WD~1856~b was an ordinary Jovian planet that underwent an extraordinary dynamical history.
\end{abstract}


\section{Introduction}
Recently, {\it TESS} observations revealed a planet-size object transiting WD~1856+534, a cool, old white dwarf (WD) with
an effective temperature of $T_{\rm eff}\simeq4700~$K
\citep{vand2020}. WD~1856~b has an orbital period of 1.4~d and a radius of $\simeq10R_\oplus$.
This orbit is so compact
that it is unlikely to have persisted throughout the red giant branch (RGB) and the asymptotic giant branch (AGB) phases of stellar evolution. Instead, WD~1856~b is thought to have formed at greater separations, and to have migrated inward after the main sequence (MS).

Although WD~1856~b is a cool ($<200$K) object, its orbit resembles those of the `hot Jupiters' that accompany ${\sim}1\%$ of MS stars \citep[e.g..][]{howard2012}. 
Among the proposed mechanisms of hot Jupiter migration, 
the von Zeipel-Lidov-Kozai\footnote{
Recently, \citet{ito2019} confirmed that
\citet{zeipel1910} carried out pioneering work on this
mechanism, predating the seminal contributions
of \citet{lid62} and \citet{koz62} by several decades.
}  (ZLK) mechanism coupled with tidal friction \citep{wu2003,FT2007} 
has emerged as a predictive and elegant contender, with
several works suggesting that, at least in part, hot Jupiters
do indeed originate from this mechanism \citep[e.g.,][]{naoz2012,petro2015,ander2016}.

Much like hot Jupiters, WD~1856~b could have migrated to
its current position due to ZLK oscillations induced 
by a known outer companion(s) to WD~1856+534 \citep{mccook1999}.
From {\it Gaia} astrometry, \citet{vand2020} measured the outer companion G~229-20, a double M-dwarf, to orbit WD~1856+534 
at a distance of  
$\sim1500$~au with an eccentricity of $\simeq0.3$
 (see Table~\ref{table:TF}). For such a wide orbit, the timescale associated to ZLK
oscillations is, to quadrupole level of
approximation,
\begin{equation}
\tau_{\rm quad}\approx  4.37\times10^7 \left(\frac{M_{\rm WD}}{0.5 M_\odot}\right)^{1/2}\left(\frac{a_{\rm p,0}}{5{\rm au}}\right)^{-3/2}\text{yr}
\end{equation}
which is much shorter than
the cooling age of the WD, estimated
to be ${\sim}6\times10^9$~yr \citep{vand2020}. 

\begin{deluxetable}{l|l}
  \tablecaption{Parameters of WD~1856+534 system \label{table:TF}}
\startdata
\\
WD mass ($M_{\rm WD})$  & $0.518\pm 0.05\; M_\odot $\\
WD cooling age ($T_{\rm cool}$) & $5.85 \pm 0.5 $ Gyr\\
\hline 
planet orbital period& $1.407$ days \\
planet semi-major axis\tablenotemark{a} ($a_{\rm p}$) & $0.0204\pm 0.0012$ au\\
planet radius $(R_{\rm p})$ & $10.4\pm 1\;R_\oplus$ \\
\hline
Mass G 229-20 A ($M_A$) & $0.346 \pm 0.027\; M_\odot$\\
Mass G 229-20 B ($M_B$) & $0.331\pm0.024\; M_\odot$\\
$M_{\rm out}=M_A+M_B$ & $0.677\pm0.051\; M_\odot$\\
A-B binary semi-major axis ($a_{\rm AB})$ & $58^{+54}_{-16}$ au\\
outer semi-major axis\tablenotemark{b} ($a_{\rm out})$ & $1500^{+700}_{-240}$ au\\
outer eccentricity $e_{\rm out}$ & $0.3^{+0.19}_{-0.1}$\\
\enddata
  \tablenotetext{a}{\small 
Fit assumes a circular orbit \citep{vand2020}.}
\tablenotetext{b}{\small The outer orbit refers corresponds to a Keplerian fit to the
separation between WD~1856+534 and the center of mass of the G 229-20
A and B pair.}
  \vspace{0.05in}
\end{deluxetable}

 Despite an earlier suggestion by \citet{agol} that the 
 ZLK mechanism could produce planets in close orbits around WDs,
 most theoretical efforts in this context have instead emphasized
 how ZLK oscillations can explain WD pollution  \citep{hamers2016,petro2017,stephan2017}. In spite of this oversight, ZLK migration of gas giants around WDs is very much possible,
 with the requirement that the planets survive {\it all} prior stages of stellar evolution.

In this work, we exploit the distinguishing feature of ZLK migration around WDs that tidal dissipation {\it in the  host} is negligible. Therefore, once the planet is parked in a circular, tidally-locked orbit, it does not decay further, in contrast to hot Jupiter systems of advanced age \citep[e.g.,][]{hamer2020}. By relating the current
semi-major axis of WD~1856~b to the maximum attainable ZLK eccentricity, we  can derive the {\it initial} separation and the planet mass
that are required for WD~1856~b to have safely migrated to its current separation while avoiding tidal disruption.

\section{Eccentricity Oscillations After the Main Sequence}
While eccentricity oscillations take place for any initial inclination $i_0$ above some angle $i_{\rm crit}$ ($39.2^\circ$ if short-range forces are absent), actual migration is only possible
at high inclinations, when the maximum eccentricity $e_{\rm max}$ surpasses a critical value $e_{\rm mig}$. Above $e_{\rm mig}$, the pericenter distance is only a few solar radii, which makes tidal dissipation effective.
Likewise, when the eccentricity is above
a critical value $e_{\rm dis}>e_{\rm mig}$,
the separation at pericenter is so small
that the planet can be tidally disrupted.

After successful migration, the semi-major axis is
\begin{subequations}\label{eq:migration_condition}
\begin{align}
\label{eq:afinal}
a_{\rm p,f}\approx 2 a_{\rm p,0}(1-e_{\rm max}),
\end{align}
\end{subequations}
subject to the condition
\addtocounter{equation}{-1}
\begin{subequations}
\setcounter{equation}{+1}
\begin{align}\label{eq:boundaries}
e_{\rm dis}{>}e_{\rm max}{\geq} e_{\rm mig},
\end{align}
\end{subequations}
where $a_{\rm p,0}$ is the planet's initial semi-major axis.
Equation~(\ref{eq:afinal}) suggests that, if $a_{\rm p,f}$ is
known ($0.0204$~au in the case of WD~1856~b; Table~\ref{table:TF}), then the planet's original orbital separation can be inferred, provided that we can calculate
if $e_{\rm max}$ terms of $a_{\rm p,0}$ and
other parameters of the system. 
\subsection{Maximum Eccentricity}
The maximum eccentricity $e_{\rm max}(i_0,\varepsilon_{\rm GR},\varepsilon_{\rm Tide})$ attainable through
quadrupole-order ZLK oscillations
satisfies the transcendental equation
\begin{equation}\label{eq:emax}
\begin{split}
\frac{9}{8}\frac{1-j_{\rm min}^2}{j_{\rm min}^2}
\bigg(\!j_{\rm min}^2-&\frac{5}{3}\cos^2\!i_0\!\bigg)
=\varepsilon_{\rm GR}\bigg(\frac{1}{j_{\rm min}}-1\bigg)\\
+&\frac{\varepsilon_{\rm tide}}{15}\bigg(\frac{1+3e^2_{\rm max}+\frac{3}{8}e^4_{\rm max}}{j_{\rm min}^9}-1\bigg)
~,
\end{split}
\end{equation}
\citep[eq. 50 of][]{liu15} where $j_{\rm min}\equiv\sqrt{1-e_{\rm max}^2}$ and 
\begin{align}\label{eq:eps_gr}
\varepsilon_\mathrm{GR}&\equiv\frac{3\mathcal{G} M_{\rm WD}^2a_{\rm out}^3(1-e_{\rm out}^2)^{3/2}}{a_{\rm p,0}^4c^2M_{\rm out}}~\\
\varepsilon_\mathrm{Tide}&\equiv\frac{15M_{\rm WD}^2a_{\rm out}^3(1-e_{\rm out}^2)^{3/2}k_{2p}R_{\rm p}^5}{a_{\rm p,0}^8M_{\rm p} M_{\rm out}}~
\end{align}
with  $k_{\rm 2p}=0.37$ being the tidal Love number of a gas giant. The coefficients $\varepsilon_\mathrm{GR}$
and $\varepsilon_\mathrm{Tide}$ represent the strength of the short-range forces --general relativistic (GR) precession and tides on the planet respectively--
relative to the tidal forcing by the external companion \citep[see also][]{FT2007}. These coefficients are fully determined from the system parameters (Table~\ref{table:TF}), except for the values of 
$i_0$, $a_{\rm p,0}$ and  $M_{\rm p}$.

\begin{figure*}[ht!]
\centering
    \includegraphics[width=0.43\textwidth]{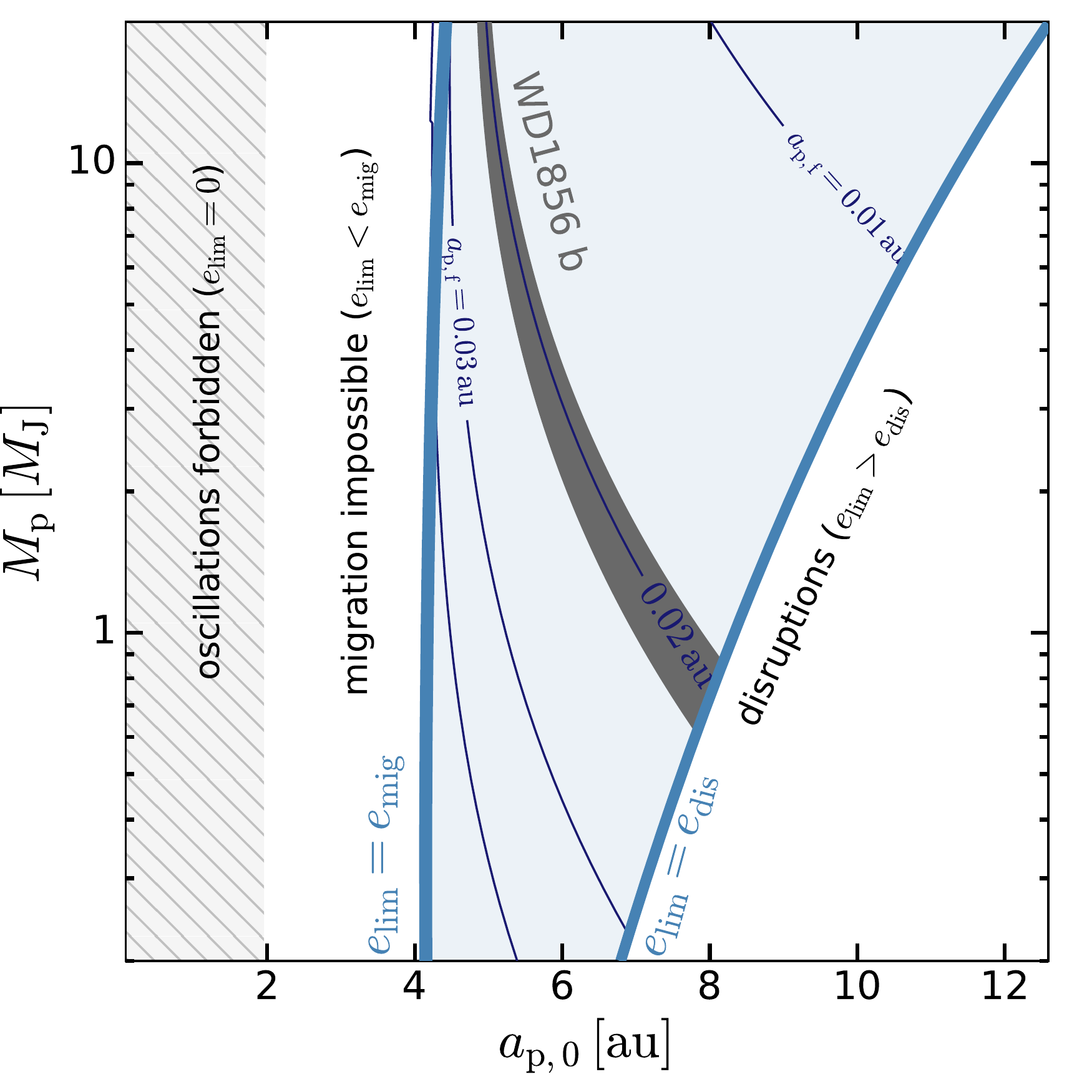}
     \includegraphics[width=0.49\textwidth]{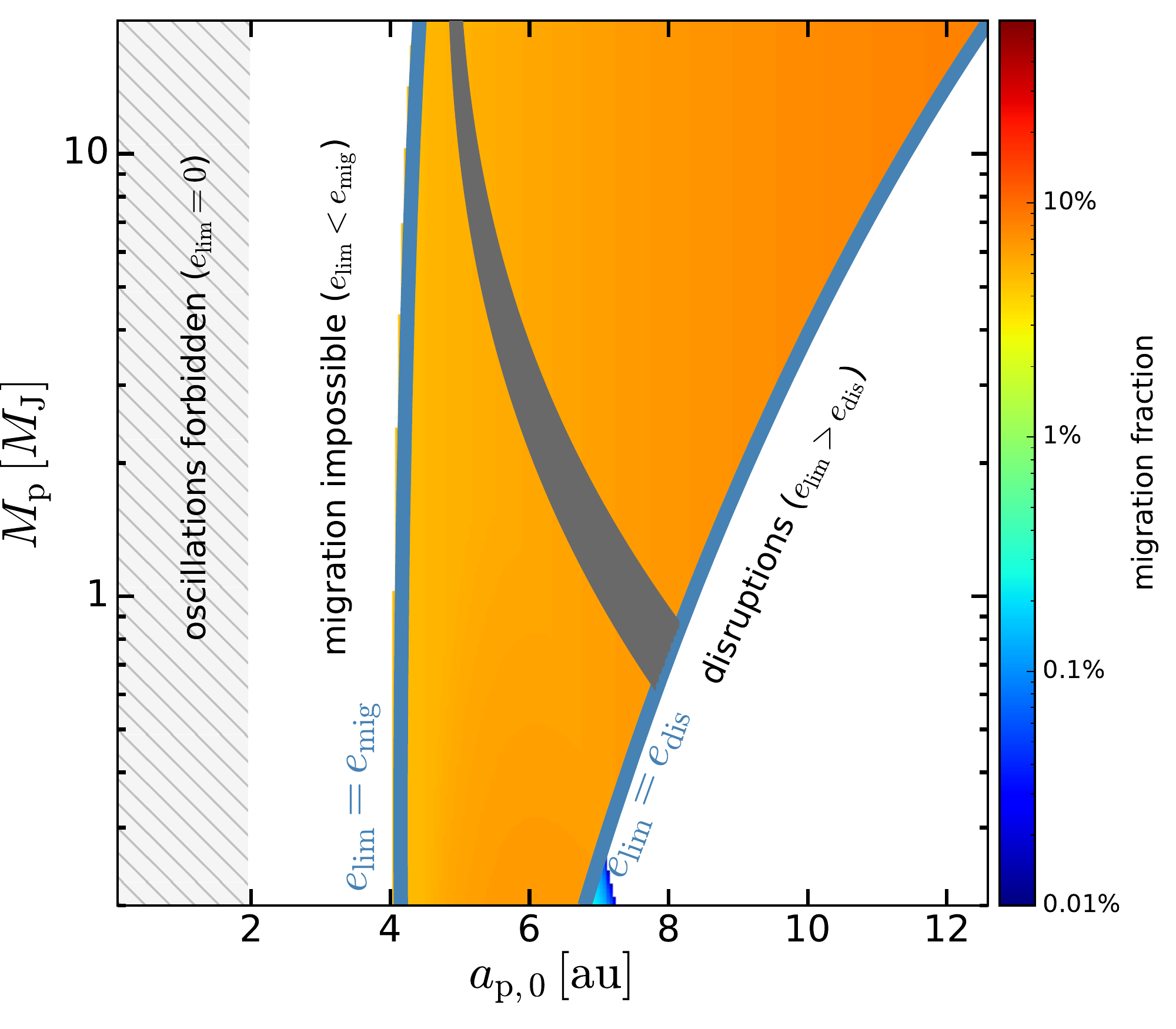}
    \caption{Kozai migration paths in the WD~1856+534 system.
    Left panel: regions of viable migration with thin dark blue contours depicting $a_{\rm p,f}$ (Equation~\ref{eq:afinal}) and thin light blue
    contours depicting the boundaries of migration ($e_{\rm lim}=e_{\rm mig}$, left) and disruption ($e_{\rm lim}=e_{\rm diss}$, right) (Equation~\ref{eq:boundaries}). The cross-hatched region corresponds
    complete quenching of ZLK oscillations owing to GR ($\varepsilon_{\rm GR}\geq9/8$; Equation~\ref{eq:eps_gr}) and the dark gray region depicts
    the current semi-major axis of WD~1856b (Table~\ref{table:TF}).
    Right panel: Kozai migration fractions calculated by the analytical method of \citet{munoz2016}. Within the migration boundaries, the migration rate is $\sim5\%$, and is primarily given by the with of the octupole window (Equation~\ref{eq:oct_window}), except for a small
    region at small planet mass and $a_{\rm p,0}\simeq7$, where the quadrupole window can also lead to migrations.
    \label{fig:migration_rates}}
\end{figure*}

At the quadrupole-level of approximation, the value of $e_{\rm max}$ grows monotonically with
$i_0$ until it reaches its upper bound, or ``limiting eccentricity'' $e_{\rm lim}$, when $\cos i_0=0$. Typically, $e_{\rm max}$ will not surpass the critical value $e_{\rm mig}$ unless
$i_0$ is very close to $90^\circ$. This narrow 
range of inclinations, $[90^\circ{-}\Delta i_{\rm quad},90^\circ{+}\Delta i_{\rm quad}]$, 
subtends a solid angle $\sin\Delta i_{\rm quad}$ that equals
the fraction of orbital orientations in the unit sphere that lead to migrations/disruptions  \citep{munoz2016}.

At the octupole-level of approximation, however,
a wider range of initial inclinations can reach extreme eccentricities \citep{katz2011},
with all angles within the ``octupole window''
$[90^\circ{-}\Delta i_{\rm quad},90^\circ{+}\Delta i_{\rm quad}]$
reaching $e_{\rm max}\approx e_{\rm lim}$ \citep{liu15}.
The width of the window is
\begin{equation}\label{eq:oct_window}
\Delta i_{\rm oct}\approx 2.9^\circ \left(\frac{\varepsilon_{\rm oct}}{10^{-3}}\right)^{1/2}
\end{equation}
\citep{munoz2016} where
$\varepsilon_{\rm oct}=(a_{\rm p,0}/a_{\rm out})e_{\rm out}(1-e_{\rm out}^2)^{-1}$
is the octupole strength parameter \citep[e.g.][]{ford2000,lithwick2011,naoz2016}, which vanishes
for non-eccentric outer companions.
For
the system WD 1856+534/G~220-20~AB, we have  
$\varepsilon_{\rm oct}\approx 1.1\times10^{-3}(a_{\rm p,0}/ 5{\rm au})$. While small, this value
of $\varepsilon_{\rm oct}$ is large enough to
provide an octupole window of $\simeq3^\circ$,
which covers a solid angle of 
$\simeq0.05$,  implying that about
$5\%$ of planets will undergo extreme eccentricity
excursions.

In most cases of interest,
the octupole window is wider than its quadrupole counterpart, which allows us to replace
$e_{\rm max}(a_{\rm p,0},M_{\rm p},i_0)$ with $e_{\rm lim}(a_{\rm p,0},M_{\rm p})$ in Equations~(\ref{eq:migration_condition}),
effectively relegating $i_0$ to a secondary role, provided
that the system is inside the octupole window.
By making this simplification, we reduce the number of unknowns to only two. For each $(a_{\rm p,0},M_{\rm p})$ pair,
we can compute $e_{\rm lim}$, and then derive a unique
value of $a_{\rm p,f}\approx2 a_{\rm p}[1-e_{\rm lim}(a_{\rm p,0},M_{\rm p})]$. In turn,
$e_{\rm lim}$ is solved from:
\begin{equation}\label{eq:elim}
\begin{split}
0=&\frac{9}{8}{e_{\rm lim}^2}
-\varepsilon_{\rm GR}\bigg[\frac{1}{(1-e_{\rm lim}^2)^{1/2}}-1\bigg]\\
&-\frac{\varepsilon_{\rm tide}}{15}\bigg[\frac{1+3e^2_{\rm lim}+\frac{3}{8}e^4_{\rm lim}}{(1-e_{\rm lim}^2)^{9/2}}-1\bigg]
~.
\end{split}
\end{equation}
We illustrate
this calculation in Figure~\ref{fig:migration_rates} (left panel), where the dark blue contours show levels of constant $a_{\rm p,f}$ for different values of
of $a_{\rm p,0}$ and $M_{\rm p}$. The tight constraints
imposed on $a_{\rm p}$ by \citet{vand2020} (gray band) translate
into a tight correlation between the values of $a_{\rm p,0}$ and $M_{\rm p}$ that can explain this system.

\subsection{High-e Migration within a Cooling Time}
The migration condition (\ref{eq:boundaries}) requires
explicit definitions of $e_{\rm mig}$ and $e_{\rm dis}$
\citep[e.g.,][]{munoz2016}. The first of these comes
from the requirement that migration must be completed on timescales shorter
than the cooling age of the WD. Thus, we require
 $\tau_{\rm dec}^{\rm (ST)}\lesssim T_{\rm cool}$,
 where 
\begin{equation}
\tau_{\rm dec}^{\rm (ST)}
=\frac{0.357}{k_{2p} \Delta t_L} \frac{M_{\rm p}}{\mathcal{G}M_*^2}\frac{a_0^8~(1-e_{\rm max})^{7}} {R_{\rm p}^5}
\end{equation}
is the
orbital decay timescale due to high-eccentricity excursions
\citep[e.g.][]{ander2016} and where
`ST' stands for the `standard tides' of weak friction
theory \citep[e.g.][]{alex1973,hut1981}. Solving for 
the eccentricity,  we obtain the minimum eccentricity required for migration
\begin{equation}\label{eq:mig}
e_{\rm mig}\equiv 
1-1.96\left(
\frac{k_{2{\rm p}}\Delta t_L T_{\rm cool}}{P_{\rm p,0}^2}\frac{M_{\rm WD}}{M_{\rm p}}\frac{R_{\rm p}^5}{a_{\rm p,0}^5}\right)^{1/7}~.
\end{equation}
Similarly, we define an eccentricity above which
disruption takes place
\begin{equation}
e_{\rm dis}\equiv 
1-\eta_{\rm dis}\frac{R_{\rm p}}{a_{\rm p,0}}\left(\frac{M_{\rm WD}}{M_{\rm p}}\right)^{1/3}
\end{equation}
where $\eta_{\rm dis}=2.7$ \citep{guillo2011}.

The two limits, $e_{\rm lim}=e_{\rm mig}$ and
$e_{\rm lim}=e_{\rm dis}$, are overlaid into
Figure~\ref{fig:migration_rates} (left panel) as light
blue curves. Migration is only possible these  boundaries. This additional requirement further constraints the mass
and original semimajor of the WD~1856~b:     $M\gtrsim0.7M_{\rm J}$ and $5~\text{au}\lesssim a_{\rm p,0}\lesssim 8~\text{au}$.

The slope of the migration and disruption limits
can be understood analytically. When $e_{\rm lim}\approx1$,
Equation~(\ref{eq:emax}) can be simplified further (see eq. 57 of \citealp{liu15}), which allows us to define  `tide-dominated'
and `GR-dominated' limits to $e_{\rm lim}$. Thus, when approaching the tidal disruption limit $1-e_{\rm lim}^2\approx({7\varepsilon_{\rm Tide}}/{27})^{2/9}$,
for which $e_{\rm lim}=e_{\rm dis}$ implies
\begin{equation}\label{eq:dis_boundary}
\begin{split}
&a_{\rm p,0}^{\rm (dis)} \simeq 8.4~{\rm au}~
\left[\frac{a_{\rm out}\sqrt{1-e_{\rm out}^2)}}{1500~{\rm au}}\right]^{6/7}
\left[\frac{\eta_{\rm dis}}{2.7} \right]^{-9/7}
\left[\frac{R_{\rm p}}{R_{\rm J}}\right]^{1/7}
\\
&\times
\left[\frac{k_{2p}}{0.37} \right]^{2/7}
\left[\frac{M_{\rm p}}{M_{\rm J}} \right]^{1/7}
\left[\frac{ M_{\rm WD}}{0.5 M_\odot} \right]^{1/7}
\left[\frac{ M_{\rm out}}{0.7M_\odot} \right]^{-2/7}
\end{split}
\end{equation}
Conversely, when approaching the migration limit, $1-e_{\rm lim}^2\approx({8\varepsilon_{\rm GR}}/{9})^{2}$, and thus,
when $e_{\rm lim}=e_{\rm mig}$ we have
\begin{equation}\label{eq:mig_boundary}
\begin{split}
&a_{\rm p}^{\rm (mig)}=
4.36~{\rm au}~
(k_{2{\rm p}}\chi {\cal T})^{-1/48}
\left[\frac{a_{\rm out}\sqrt{1-e_{\rm out}^2}}{1500~{\rm au}} \right]^{7/8}
 \\
&\times
\left[\frac{R_{\rm p}}{R_{\rm J}}\right]^{-5/48}
\left[\frac{ M_{\rm WD}}{0.5 M_\odot} \right]^{13/24}
\left[\frac{ M_{\rm out}}{0.7M_\odot} \right]^{-7/24}
\left[\frac{M_{\rm p}}{M_{\rm J}} \right]^{1/48}
\end{split}
\end{equation}
where $\chi=\Delta t_L/0.1{\rm s}$ and
${\cal T}=T_{\rm cool}/1{\rm Gyr}$. From these analytical
expressions, we see that the dependence
of $a_{\rm p}^{\rm (mig)}$ and $a_{\rm p}^{\rm (dis)}$
is weak on most parameters except for $a_{\rm out}$, which
underscores the importance of having a well characterized
outer orbit when estimating the migration viability
and the migration fraction.

\paragraph{Migration Fractions}
In Figure~\ref{fig:migration_rates} (right panel), we compute
the Kozai migration fraction around WD~1856+534 following
the approximated method of \citet{munoz2016}. The figure shows
that the migration rate is vastly dominated by the octupole
window (except at $a_{\rm p,0}\sim7$~au and $M_{\rm p}\lesssim 0.2M_{\rm J}$), which results in a migration rate given by the solid
angle $\sin\Delta i_{\rm oct}\approx 5\%$ regardless of planet mass
and initial semi-major axis.

\subsubsection{Fast Migration and Chaotic Tides}\label{sec:chaotic_tides}
The approximate method laid out above implicitly assumes
that the dissipation rate is low enough such that the energy
is conserved over the ZLK timescale $\tau_{\rm quad}$, and
that dissipation does not preclude $e_{\rm lim}$ from being reached.
In principle, however, and under highly dissipative conditions, enough orbital energy can be lost during just one ZLK cycle
to decouple the planet from the companion's tidal field, halting
subsequent oscillations.
This regime --referred
to as ``fast migration''  by \citet{petro2015}--  can
cap the maximum eccentricity to some value $e_{\rm fast}^{\rm (ST)}$ 
and shield planets from being tidally disrupted if 
$e_{\rm fast}^{\rm (ST)}<e_{\rm dis}$. In most cases, however,
$e_{\rm fast}^{\rm (ST)}$ is not low enough to prevent disruption,
unless unrealistically large values of $\Delta t_L$ are used \citep{petro2015}.

\begin{figure}[t!]
\centering
    \includegraphics[width=0.46\textwidth]{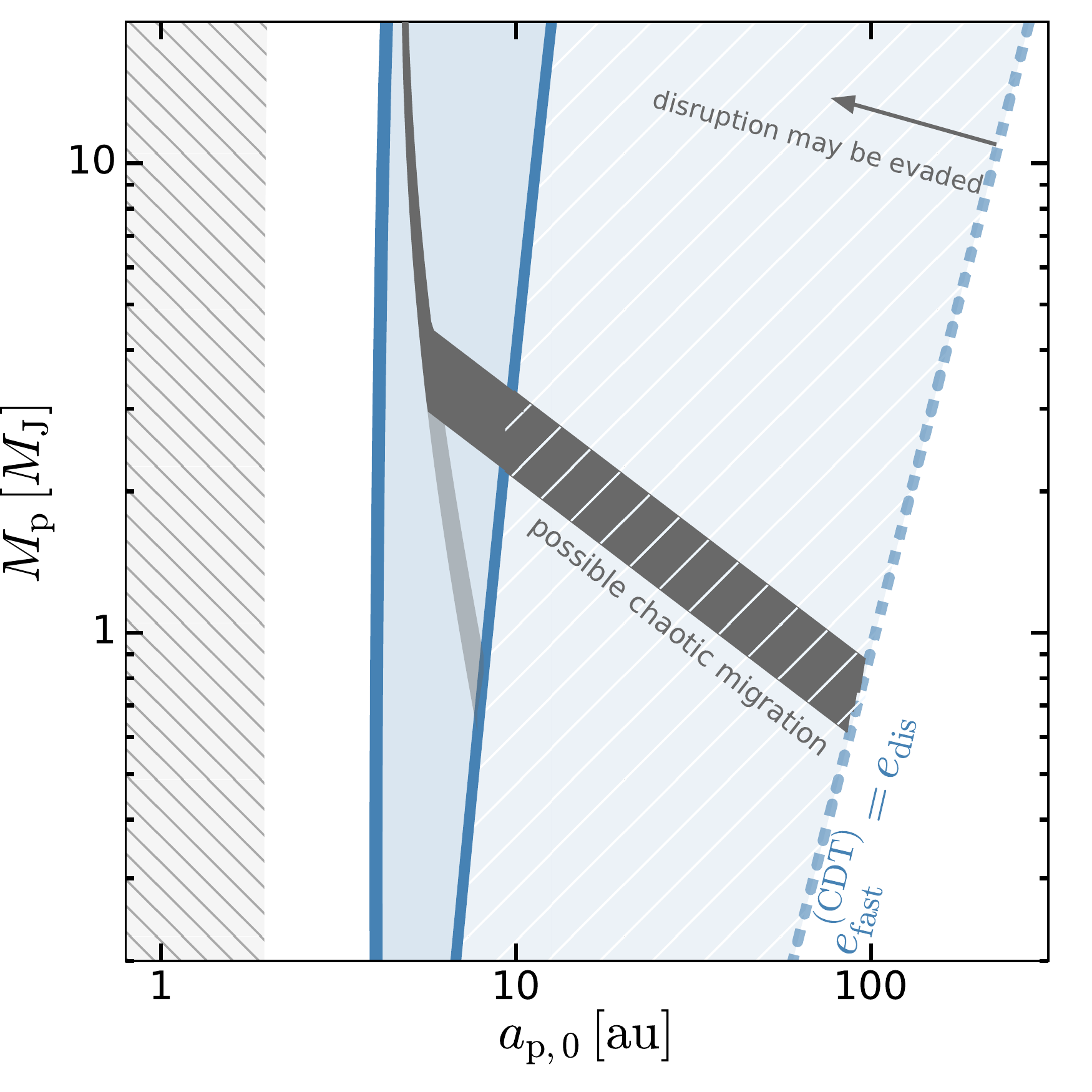}
    \caption{Same as Figure~\ref{fig:migration_rates} (left panel), but
    after replacing $e_{\rm lim}$ with $\min[e_{\rm lim},e_{\rm fast}^{\rm (CDT)}]$ (Equation~\ref{eq:e_fast}) to take chaotic tides into account 
    (Section~\ref{sec:chaotic_tides}). The cross-hatched area highlights
    the region in which disruption may be evaded by chaotic tides and migration is possible; this region is probabilistic, with both disruptions and migrations taking place in roughly equal fractions \citep{vick2019}.
    \label{fig:chaotic_tides}}
\end{figure}

An analogous, yet more efficient,
effect can be accomplished via chaotic dynamical
tides \citep[e.g.][]{mard1995,vick2018,wu2018}.
In this mechanism, the planet's fundamental
mode of oscillation is erratically excited/reduced at each pericenter passage. The mode can
grow stochastically until it ``breaks'', dissipating a significant amount of energy. 
The CDT dissipation timescale is given by 
\begin{equation}
\tau_{\rm dec}^{\rm (CDT)}
=\pi\frac{M_{\rm p}\sqrt{{\cal G}M_{\rm WD}a_{\rm p,0}}}{\Delta E_\alpha}~,
\;\;\;\;
\Delta E_\alpha = \frac{{\cal G}M_{\rm p}^2}{R_{\rm p}} \eta^{-6} T_{22}
\end{equation}
where $\eta\equiv (a_{\rm p,0}/R_{\rm p})(M_{\rm p}/M_{\rm WD})^{1/3}(1-e)$ and
$\Delta E_\alpha$ is the amount energy injected into the f-mode that is dissipated
\citep[e.g.,][]{lai1997} and $T_{22}$ is a dimensionless function \citep{press1977}. Using values
derived by \citet{vick2019} for a polytropic model of a gas,
we can approximate
$T_{22}\approx2\times10^{3}\eta^{-10}$ 
for $\eta\gtrsim\eta_{\rm dis}= 2.7$.
The requirement for fast migration, just like in the `standard tides' case,
stems from the requiring that $\tau_{\rm dec}$ is shorter
than the time spent above an eccentricity $e$ during ZLK oscillations,
i.e., $\tau_{\rm dec}\lesssim \tau_{\rm quad}\sqrt{1-e_{\rm max}^2}$
\citep[e.g.,][]{ander2016}. Solving for the eccentricity, we find
\begin{equation}\label{eq:e_fast}
\begin{split}
e_{\rm fast}^{\rm (CDT)}
\gtrsim
1-1.55\bigg[&
\left(\frac{R_{\rm p}}{a_{\rm p,0}}\right)^{15}\left(\frac{M_{\rm WD}}{M_{\rm p}}\right)^{13/3}
\frac{M_{\rm WD}}{M_{\rm out}}\\
&\times\left(\frac{a_{\rm out}}{a_{\rm p,0}}\right)^3(1-e_{\rm out}^2)^{3/2}\bigg]^{2/31}~.~~~~~~~~~~~~~~~~~~
\end{split}
\end{equation}
\begin{figure*}[t!]
\centering
    \includegraphics[width=0.48\textwidth]{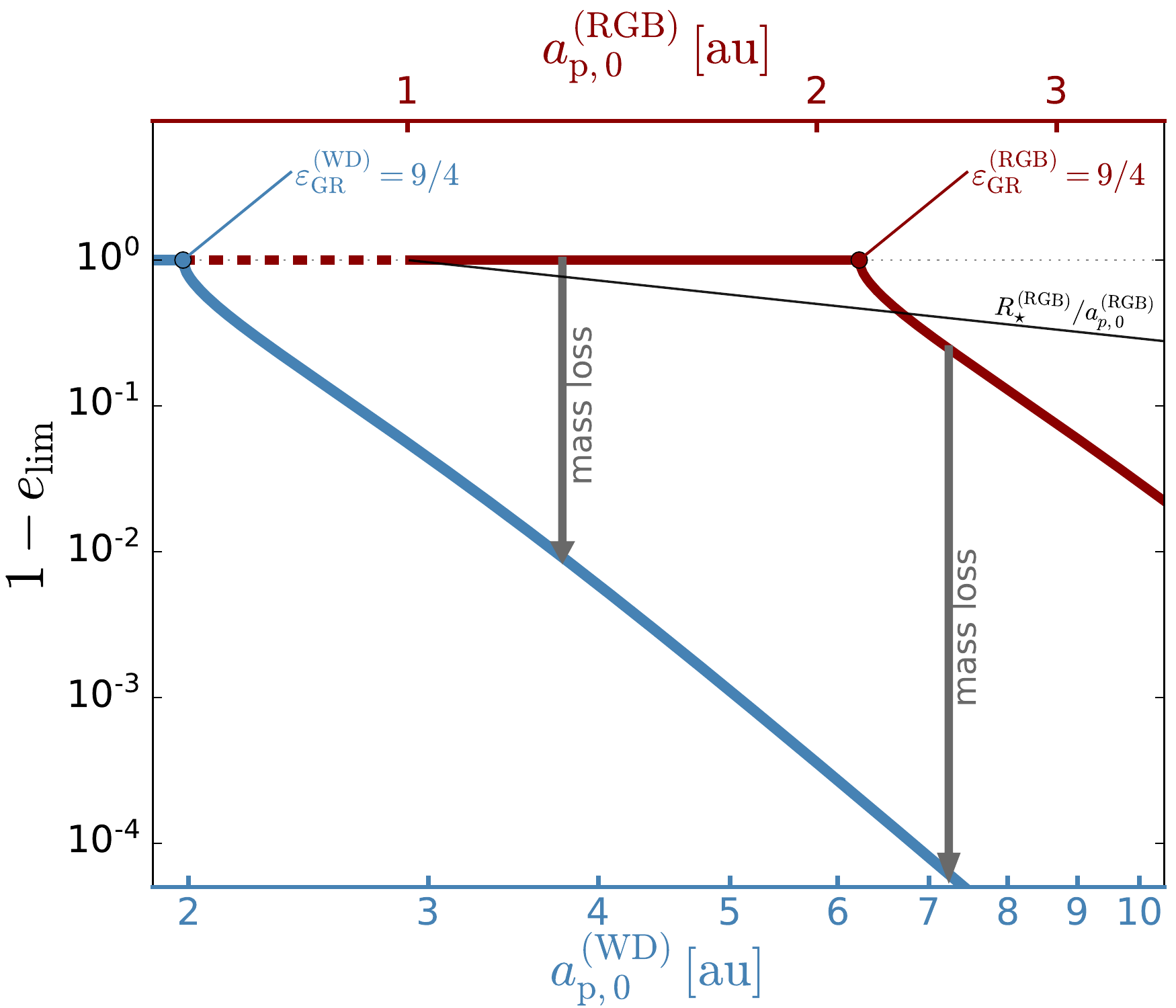}
    \hspace{0.15in}
         \includegraphics[width=0.37\textwidth]{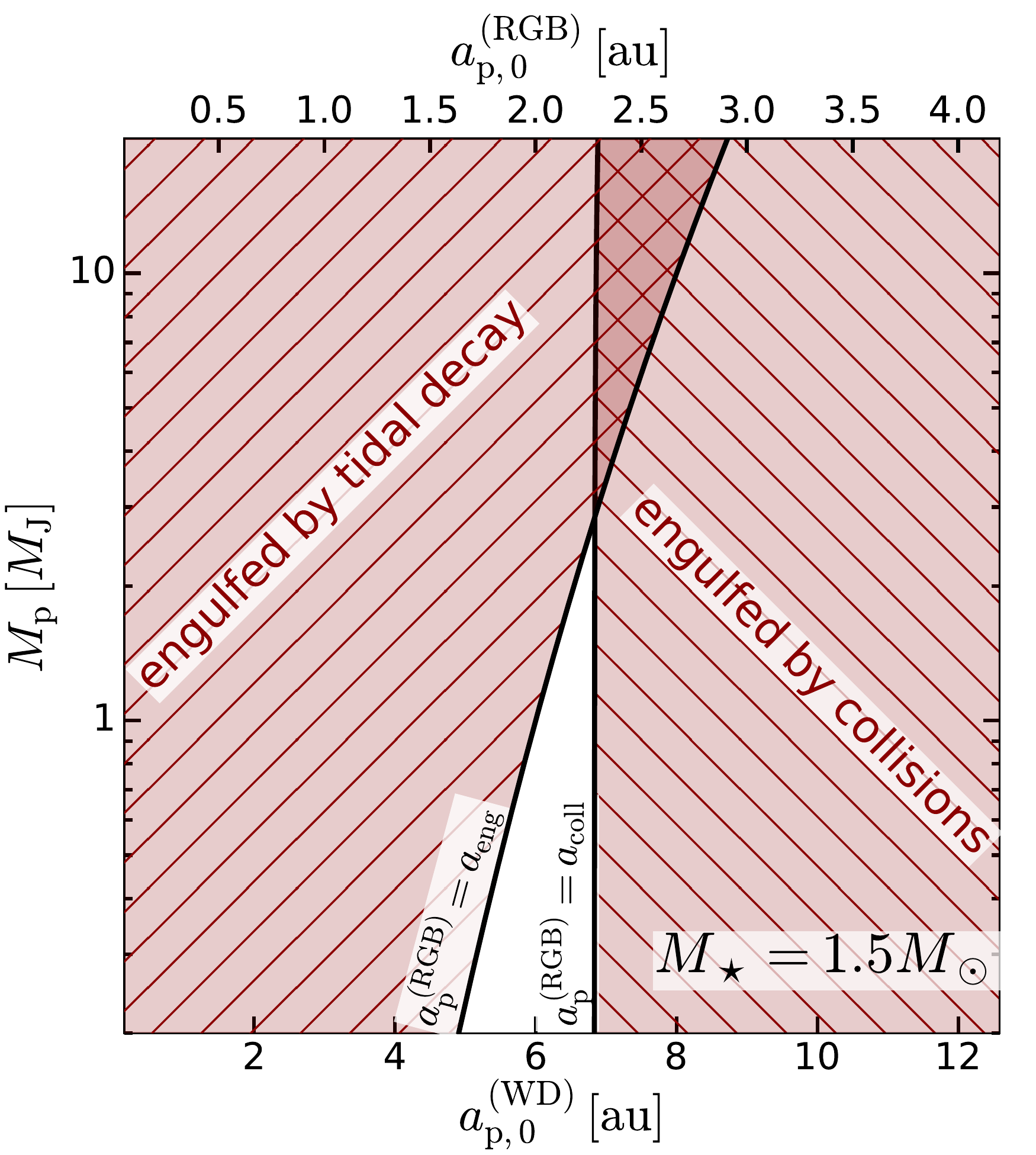}
    \caption{ZLK oscillations in the RGB phase. Left panel: modification
    of $e_{\rm lim}$ after a change in $\varepsilon_{\rm GR}$
    triggered by mass loss between the RGB (red) and WD (blue) phases (Equation~\ref{eq:eps_change}). The RGB phase
    exhibits fully quenched ZLK for $a_{\rm p,0}^{\rm (RGB)}\lesssim2.15$
    (when $\epsilon_{\rm GR}^{\rm (RGB)}=9/4$); this region is awakened
    after a mass change of $1M_\odot$, exposing all planets
    with $a_{\rm p,0}^{\rm (WD)}\gtrsim0.68$ to ZLK oscillations. For reference, we also depict $R_\star^{\rm (RGB)}/a_{\rm p,0}^{\rm (RGB)}$
    which intersec $1-e_{\rm lim}$ at 2.3~au (Equation~\ref{eq:a_coll}). Right panel: planet
    survival during the RGB phase. Semimajor axes greater than 
    $a_{\rm coll}\approx2.3$~au get too eccentric to survive collision
    with the inflated stellar envelope. Semimajor axes smaller
    than the $a_{\rm p}\propto M_{\rm p}^{1/8}$ line decay because of
    tides in the star \citep[e.g.,][]{villaver}. The white region is the only one that survives at high inclinations.
    \label{fig:gr_change}}
\end{figure*}

In Figure~\ref{fig:chaotic_tides}, we repeat the calculation leading to Figure~\ref{fig:migration_rates},
but replacing $e_{\rm lim}\rightarrow\min[e_{\rm lim},e_{\rm fast}^{\rm (CDT)}]$. The effect of chaotic tides
is readily appreciated by the dramatic shift of the disruption
boundary toward much larger initial semi-major axes\footnote{
The migration boundary on the left is also affected by chaotic
tides, producing circularized orbits at greater separations \citep{vick2019}, but this modification is of lesser
relevance for objects like WD~1856~b, which lies close to the disruption
boundary.
} 
(see fig. 10 in \citealp{vick2019}),
seemingly expanding
the parameter space of orbits that could explain WD~1856~b (cross-hatched region). Within this greatly expanded parameter space, 
the migration condition $a_{\rm p,f}=0.02$~au (Equation~\ref{eq:migration_condition}) constrains the planet mass 
within a factor of 2, but at the expense of a highly uncertain initial semi-major axis. Fortunately, the true viability of this expanded region is
severely limited if we additionally require planets to have survived earlier phases of stellar evolution. Below, we show that survival during
the RGB largely rules out the chaotic tide domain.

\section{Pre-WD phase}
Having shown that WD~1856~b could have successfully migrated via ZLK oscillations from much larger separations,
we now turn to addressing
if such a planet could have orbited a WD in the first place, having survived the MS and the subsequent giant phases.

\subsection{Quenched ZLK Oscillations Before Mass Loss}
\label{sec:mass_loss}
It is known that mass loss
can awaken ``dormant'' secular
instabilities in triples
and multiples. The driver behind
this awakening is the unequal expansion
of the orbits.
For example, in the so-called 
`mass-loss induced eccentric Kozai' (MIEK) mechanism \citep{shappee2012}, $\epsilon_{\rm oct}{\propto} a_{\rm p,0}/a_{\rm out}$ grows, and alongside it, so does the width of the octupole window (Equation~\ref{eq:oct_window}),
which can promote mild eccentricity oscillations into extreme ones.

Similarly, the expansion of the semi-major axes due to mass loss changes the balance of short-range forces in Equation~(\ref{eq:elim}). If a 
star of mass $M_\star$ loses 
an amount $\Delta M$ adiabatically,
 then the semi-major axis of the planet changes as 
$a_{\rm p}{\rightarrow} a_{\rm p}M_\star/(M_\star-\Delta M)$, while
that of the binary changes as
$a_{\rm out}{\rightarrow} a_{\rm out}(M_\star+M_{\rm out})/(M_\star{+}M_{\rm  out}{-}\Delta M)$. Consequently, the GR
coefficient changes by an amount
\begin{equation}\label{eq:eps_change}
\varepsilon_\mathrm{GR}
\rightarrow\varepsilon_\mathrm{GR}
\frac{(M_\star-\Delta M)^6}{M_\star^6}
\frac{(M_\star+M_{\rm out})^3}{(M_\star{+}M_{\rm  out}{-}\Delta M)^3}~,
\end{equation}
which modifies the moderating or quenching effect that GR exerts on ZLK oscillations.
The change in the GR coefficient is significant if we consider the mass loss rates of
G, F and A stars toward the end of their
respective AGB phases. With $\Delta M \simeq0.5-1.5M_\odot$ for WD~1856+534 \citep{cummings}, the respective change in 
$\varepsilon_{\rm GR}$ is $\simeq0.05-0.003$.
We further illustrate this effect in
Figure~\ref{fig:gr_change}, were
we depict $e_{\rm lim}$ (Equation~\ref{eq:elim}) as a function of
planet semi-major axis before and after mass loss.

\subsection{Surviving the RGB phase}
The RGB phase of stellar evolution imperils any planet orbiting at a distance of a few au, compromising  the planet's chances of surviving
all the way to the WD phase.
These planets are directly affected by the inflated stellar envelope in two ways.
First, by stellar tides: the greatly expanded star
makes it susceptible to planet-induced tides, which can shrink the orbit effectively, leading to engulfment
if
\begin{equation}\label{eq:a_engulf}
a_{\rm p}^{\rm (RGB) }\leq 
a_{\rm eng}\equiv 2 \mbox{ au }\left(\frac{M_{\rm p}}{1M_{\rm J}}\right)^{1/8}~
\end{equation}
\citep[e.g.][]{villaver}, where 
$a_{\rm p}^{\rm (RGB) }$ denotes the planet semi-major axis during the RGB phase. We caution that this boundary is fuzzy and highly dependent on the tidal model.

And second, by direct high-eccentricity collisions:
in the presence of a binary companion, ZLK oscillations
can lead to the planet being engulfed by directly plunging it into the stellar envelope; this condition reads
\begin{equation}\label{eq:a_coll}
a_{\rm p}^{\rm (RGB)}
\leq a_{\rm coll}\equiv R_{\star}^{\rm(RGB)}(1-e_{\rm lim}^{\rm (RGB)})^{-1}
\end{equation}
where $ R_{\star}^{\rm(RGB)}\simeq1$~au is the maximum
radius reached by the stellar envelope in the RGB phase
\citep[e.g.,][]{villaver}, and where
$e_{\rm lim}^{\rm (RGB)}$ is the solution to 
Equation~(\ref{eq:elim}) evaluated with parameters appropriate for the RGB phase at peak radius\footnote{
For simplicity, we assume that the stellar mass 
at the RGB and MS phases is the same. For a more detailed
modeling of mass loss coupled to ZLK oscillations, see \citet{stephan2018}.
}. 

The joint requirement $a_{\rm eng}< a_{\rm p}^{\rm (RGB)}<a_{\rm coll}$ guarantees survival of planets throughout the RGB phase.  As
it turns out, this condition is difficult to satisfy, and a large fraction of $a_{\rm p}^{\rm (RGB)}{-}M_{\rm p}$
space is excluded (Figure~\ref{fig:gr_change}, right panel).
The engulfment condition (Equation~\ref{eq:a_engulf}) and the collision condition (Equation~\ref{eq:a_coll}) 
conspire to create a narrow region (in white) that allows inclined planets to survive.

\begin{figure*}[t!]
         \includegraphics[width=0.325\textwidth]{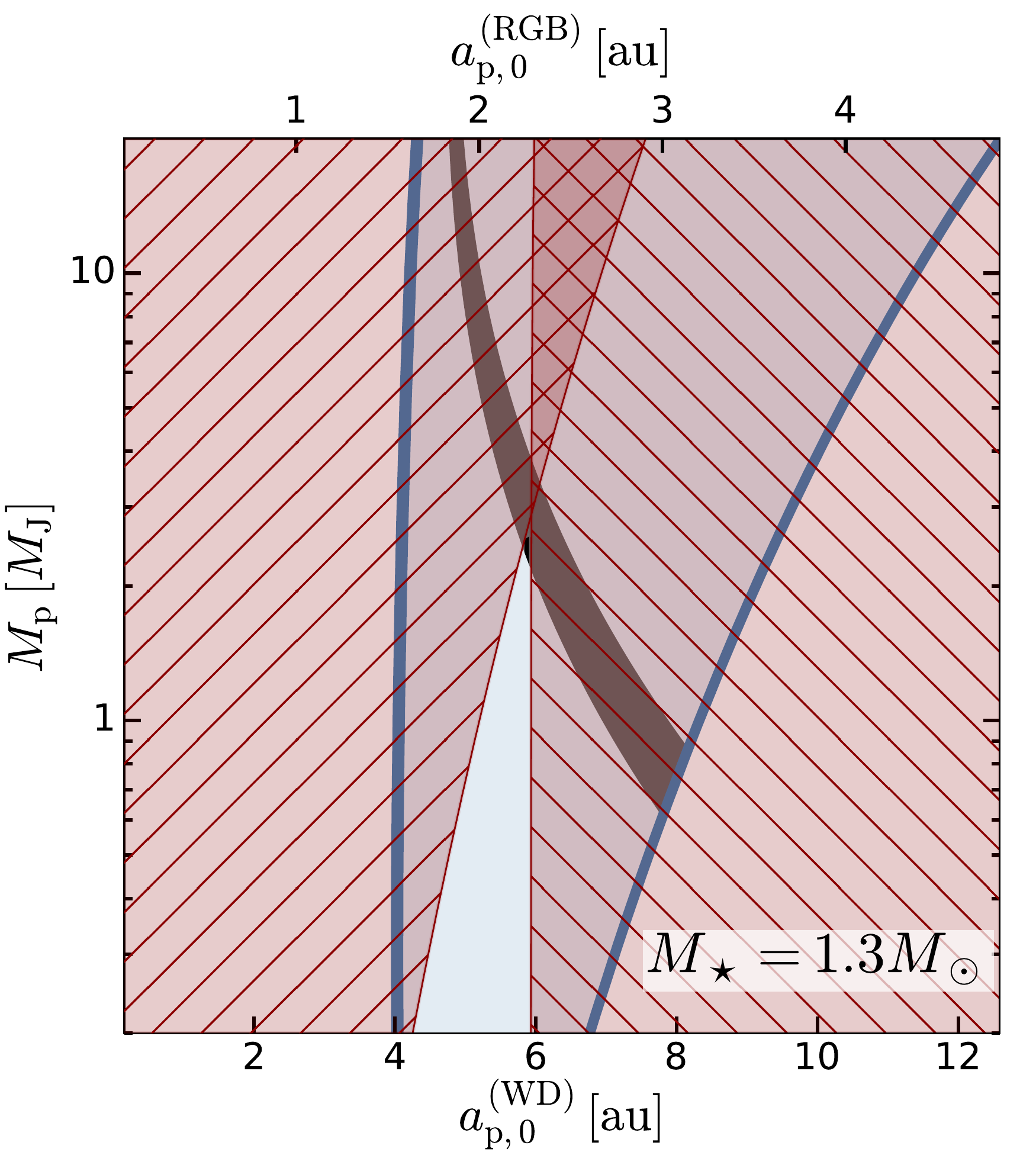}
     \includegraphics[width=0.325\textwidth]{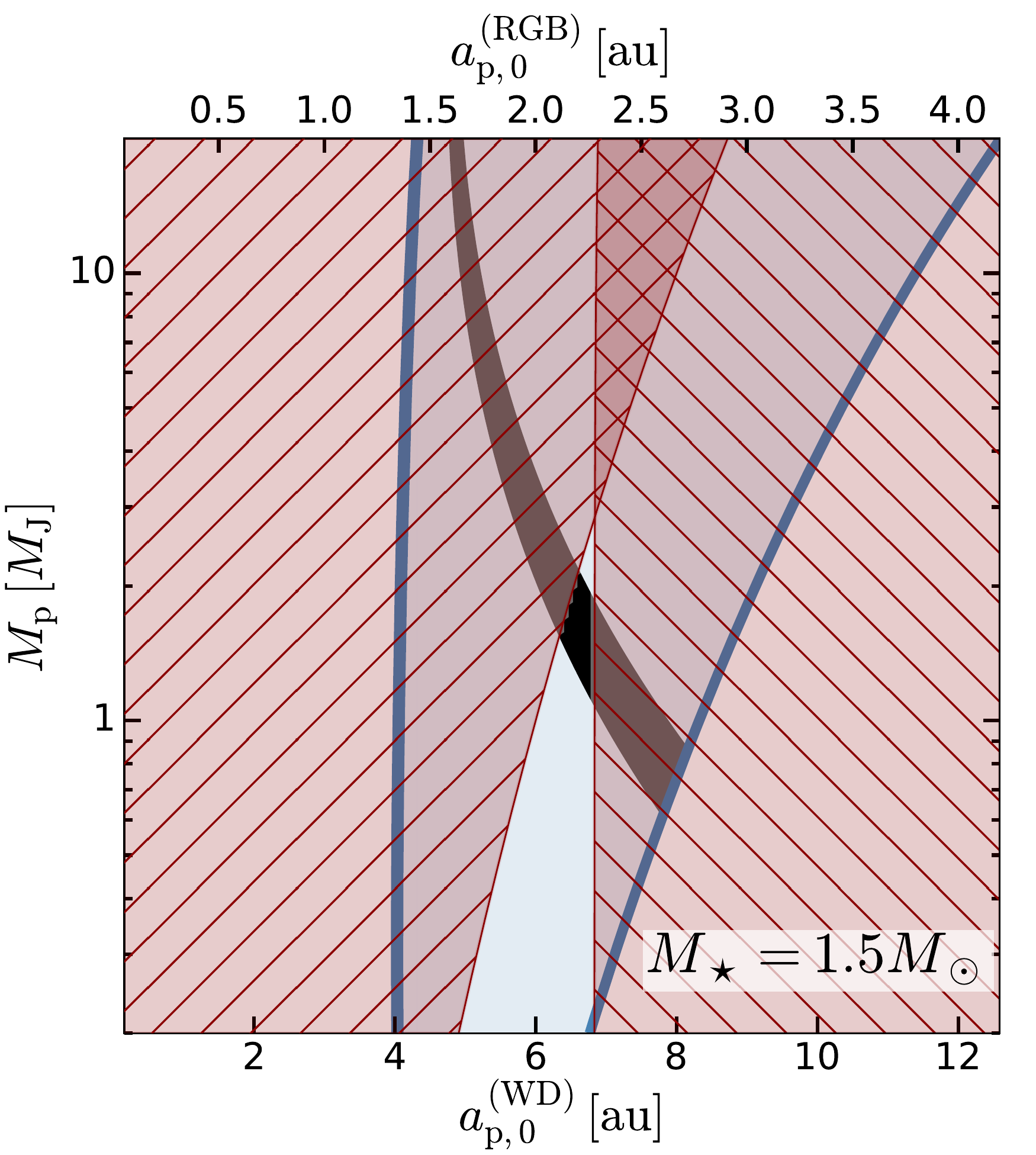}
      \includegraphics[width=0.325\textwidth]{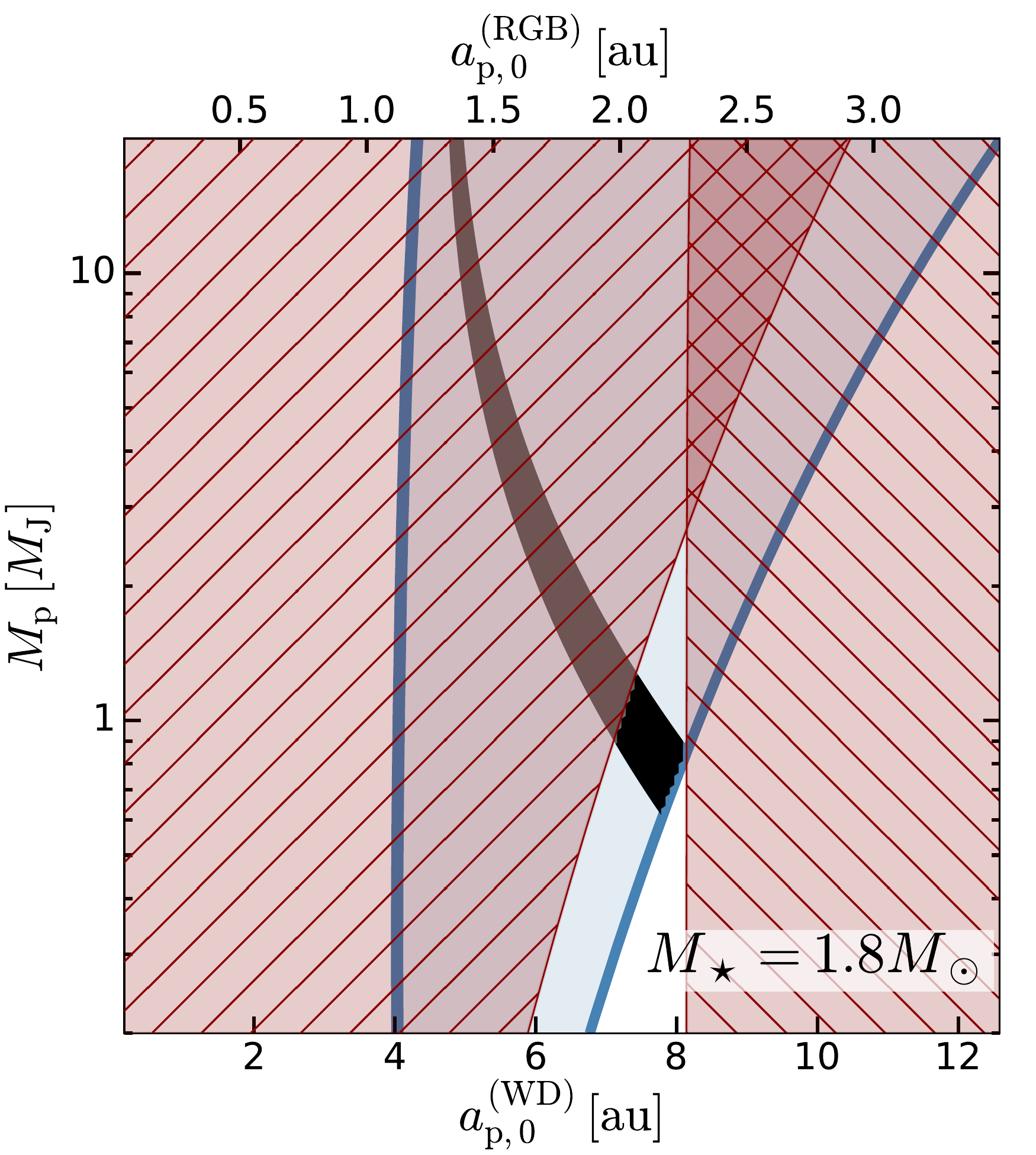}
    \caption{Survival and migration viability of highly inclined planets  throughout stellar evolution. We combine the requirements Figures~\ref{fig:migration_rates}
    and~\ref{fig:gr_change} (right panel) to constrain the initial
    conditions of WD~1856~b for different values of the 
    initial stellar mass $M_\star$. The F star (left panel) does not
    survival and migration; the late A star (middle panel) allows for survival and migration via standard tides, the early A star (right panel) may allow for survival and migration via chaotic tides.
    \label{fig:planet_survival}}
\end{figure*}

We are now in a position to combine the migration viability conditions
of Figure~\ref{fig:migration_rates} with the survival
conditions of  Figure~\ref{fig:gr_change} (right panel),
to identify which regions of parameter space give WD~1856~b a viable path to its current 
orbit. We show these overlaid conditions in Figure~\ref{fig:planet_survival} for different values
of $M_\star$ --the initial stellar mass. This figure
shows that if $M_\star\leq1.3M_\odot$ (left panel) there
is no possible path for WD~1856~b to have survived and migrated.
Conversely, there is a narrow range of parameters that 
explains WD~1856~b's current orbit if $M_\star\simeq 1.5-1.8~M_\odot$ (middle and right panels), which roughly corresponds to
$M_{\rm p}\simeq0.7-3M_{\rm J}$ and $a_{\rm p,0}=2-2.5a_{\rm p,0}$.

At even higher initial stellar masses, the survival region overlaps with that dominated by disruptions and/or chaotic tides (Section~\ref{sec:chaotic_tides}). Initial masses
above $1.8M_\odot$, however, are unlikely to produce WD~1856+534 \citep{cummings}, 
indicating that chaotic tides play a minor role, if any, in enabling
planets with $a_{\rm p,0}\sim 2.5$~au to be precursors of WD~1856~b.

\section{Discussion}
We have demonstrated that Kozai migration can explain the recent discovery of the planet-size companion to WD~1856+534. By simultaneously requiring that the companion is not lost prematurely during the RGB phase and that it subsequently migrates to its current location, we are able to constrain the planet's
initial semi-major axis ($\simeq 2-2.5$~au)
and mass ($\simeq 0.7-3 M_{\rm J}$).

\subsection{Occurrence rate}
To provide with an estimate
of the occurrence rate of WD-transiting Jovians, we
proceed as follows. We assume that the progenitor is always an A star. The binarity fraction of A-type stars is ${\cal F}_{\rm bin}\simeq 0.7$ \citep{derosa} and their giant planet-bearing fraction is ${\cal F}_{\rm Jup}\simeq0.2$ \citep{occurrence}. We assume that the semi-major axes $a_{\rm p}$ and $a_{\rm out}$, and the mutual inclination $i_0$ follow independent distributions. We then define the fraction of systems that
survive stellar evolution and subsequently migrate into a close-in orbit as
\ba\label{eq:f_mig}
&&{\cal F}_{\rm surv, mig}=
\int
d\log a_p d\log a_{\rm out}d\cos i_0
\frac{dN}{d\log a_{\rm p}}
\frac{dN}{d\log a_{\rm out}}\nonumber\\
&\times&\Theta\left[a_{\rm p}-2\mbox{ au}\right]\times\Theta\left[(1-e_{\rm lim}^{\rm (RGB)})-R_\star^{\rm (RGB)}/a_p\right]\nonumber\\
&\times&
\Theta\left[2R_\odot/a_p^{\rm (WD)}- (1-e_{\rm lim}^{\rm (WD)})\right]
\times \Theta[2.6\varepsilon_{\rm oct}^{\rm (WD)}-\cos^2i_0],\nonumber\\
\ea
where $\Theta$ is the Heaviside function. The last term in the integral represents the effects of the octupole window (Equation~\ref{eq:oct_window}).

We evaluate ${\cal F}_{\rm surv, mig}$ assuming log-normal 
distributions in the semi-major axes:  $\ln (a_{\rm p,0}/{\rm au})\sim {\cal N}(0.92,0.7)$ 
and $\ln (a_{\rm out}/{\rm au})\sim {\cal N}(5.97,0.78)$, in broad agreement with \citet{fernandes}  and \citet{derosa}, respectively. For simplicity, we assume that all wide binary companions have $e_{\rm out}=2/3$\footnote{
Mean value for a thermal distribution, which is appropriate for wide binaries.}. 
This yields ${\cal F}_{\rm surv, mig}\simeq 0.3\%$. 

The net fraction of WDs hosting close-in Jovians owing to Kozai migration is
\ba\label{eq:f_jup}
{\cal F}_{\rm WD,Jup}&=& {\cal F}_{\rm Jup}\times {\cal F}_{\rm b}\times {\cal F}_{\rm surv,mig}\nonumber\\
&\simeq& 0.2\times0.7\times 0.003\simeq 4\times10^{-4}
\ea
or one planet per $\simeq2,500$ WDs, which is consistent with the $3-\sigma$ upper limits of $0.45\%$ derived from photometric surveys \citep{fulton,eylen}. However, we cannot rule out other migration mechanisms
that could increase ${\cal F}_{\rm WD,Jup}$. For example, the gas giant around WD 1145+017 \citep{gasincke} is too young ($\sim13$ Myr), and its orbital separation too wide ($\sim0.07$ au), to be explained by Kozai migration \citep{VF2020}.

Further constraints on the occurrence of close-in Jupiters around WDs are expected in the near future from various surveys, including  LSST with expected yields of $10^7$ surveyed WDs  \citep{agol}. With a detectability of $\simeq2\%$ for WD 1856b-like planets \citep{CK2019}, we expect $\sim100$ Kozai-migrated planets to be discovered in the the ten year
baseline of LSST.

\subsection{Increasing Occurrence with Additional Effects}
The main bottleneck in the low occurrence rate
is the joint requirement of past survival plus a late-onset migration, which severely restricts the viable region of 
parameter space (Figure~\ref{fig:planet_survival}).
These calculations, however, are sensitive to
the choices of $a_{\rm eng}$ and $a_{\rm coll}$ 
(Equations~\ref{eq:a_engulf} and~\ref{eq:a_coll}), as well
as the disruption boundary (Equation~\ref{eq:dis_boundary}),
and are thus subject to caveats. 

A way of expanding the survival window (Figure~\ref{fig:gr_change}, right panel) is to incorporate
additional sources of apsidal precession to shift the $a_{\rm p}^{\rm (RGB)}=a_{\rm coll}$ boundary to the right.
This effect can be accomplished by adding a planetary system
interior to $\sim2$ au, as proposed by \citet{petro2017}.
Additional planets quench ZLK oscillations, which can be
triggered once the planets are engulfed in the RGB or AGB phases (see \citealt{ronco} for engulfment in multi-planet systems).

\subsection{Relation to Previous Work}
In this work, we have used mass loss
to trigger an otherwise suppressed ZLK mechanism. Delaying the onset
of ZLK oscillations was instrumental for constraining the past and present properties of WD~1856~b.
More generally, though, mass loss can trigger varied responses
\citep[e.g.,][]{KP12,veras2013}, and it can lead to dynamical and secular instabilities
that are effective at transporting material/minor bodies toward the WD when binaries are present (see \citealt{veras_review} for a review).

 One such possibility is the enhanced effect of galactic tides for very wide binaries ($\gtrsim 5000$ au). The galactic tide that can make $e_{\rm out}$ grow within a cooling age, thus disturbing a planetary system \citep{bonsor}. In the case of
 WD 1856+534, however, the outer companions are too close for the galactic tide to operate. 

A second possibility is the MIEK mechanism discussed above,
which produces the widening the "octupole window",
promoting conventional (quadrupolar) ZLK oscillations into extreme (octupolar) ones \citep{shappee2012,hamers2016,stephan2017}. 
But the applicability of this mechanism is limited
in the case of WD 1856+534, because the octupole
window is known to be narrow ($\simeq3^\circ$), even after being widened by mass loss. If planets were to be promoted
into octupolar oscillations, it would be from already large inclinations,
which would accompanied by large amplitude (quadrupolar) oscillations during the MS and RGB phases. Consequently, in order to survive the RGB phase, these planets would need to start from very large initial semi-major axes
($\sim $100~au), like in the simulations of
 \citet{hamers2016} and \citet{stephan2017}. The scarcity of Jovians planets at such large distances from the host star  \citep{fernandes} renders this type of mechanism improbable.

\subsubsection{Dynamics of 2+2 Systems}
We have treated the outer M-dwarf binary as single body of mass $M_{\rm out}=M_A+M_B$. This approximation may break-down in some regimes, especially when  the quadrupolar field from the double M-dwarf modulates the wide binary on timescales comparable to $\tau_{\rm quad}$. Under certain conditions, the quadruple system can evolve
chaotically, with the eccentricity diffusively evolving toward extreme values \citep{HL2017}. In the WD~1856+534 system and $a_{\rm p,0}\simeq5-8$ au, however, the dimensionless quantity
\vspace{0.05in}
\begin{equation}\label{eq:chaotic_coeff}
\frac{3}{4}\left(\frac{M_{\rm WD}}{M_{\rm out}}\right)^{3/2}\left(\frac{a_{\rm AB}}{a_p}\right)^{3/2}
\sim 10-28
\end{equation}
is too large for this chaotic diffusion to operate, and we can
thus safely treat the system as a triple. It is nonetheless
worthy of mention that mass loss can increase Equation~(\ref{eq:chaotic_coeff}), and
conceivably activate the chaotic 2+2 dynamics for some systems with larger semi-major axes ($a_{\rm p,0}\sim a_{\rm AB}$) after a WD is formed.

\section{Conclusion}
We have shown that
the current orbit of WD~1856~b can be explained with
Kozai migration without any ad hoc requirements other than a highly inclined orbit respect to the outer companions. By requiring that WD~1856~b survived stellar evolution and that its migration began during the WD phase, we are able to constrain its initial semi-major axis and mass.
We infer an initial separation of $2-2.5$~au, and a mass of
$0.7-3M_{\rm J}$, implying that WD~1856~b was born a typical gas giant.

Although the initial conditions we have derived are typical of planetary systems, the planets that survive till the end of the WD phase are rare. 
We predict the occurrence rate of close-in Jovians from Kozai migration
around WDs to be ${\cal O}(10^{-3}{-}10^{-4})$ and expect that LSST will find ${\sim}100$ of such systems.

\acknowledgments{We are grateful to Andrew Youdin, Kaitlin Kratter, Mar\'ia Paula Ronco, and Max Moe for useful discussions. DJM acknowledges support from the CIERA Fellowship at Northwestern University and the Cottrell
Fellowship Award from the Research Corporation for Science Advancement
which is partially funded by the NSF grant CHE-2039044.
 CP acknowledges support from the Bart J. Bok fellowship at Steward Observatory.}


\bibliographystyle{aasjournal}

\begin{thebibliography}{}
\expandafter\ifx\csname natexlab\endcsname\relax\def\natexlab#1{#1}\fi
\providecommand{\url}[1]{\href{#1}{#1}}
\providecommand{\dodoi}[1]{doi:~\href{http://doi.org/#1}{\nolinkurl{#1}}}
\providecommand{\doeprint}[1]{\href{http://ascl.net/#1}{\nolinkurl{http://ascl.net/#1}}}
\providecommand{\doarXiv}[1]{\href{https://arxiv.org/abs/#1}{\nolinkurl{https://arxiv.org/abs/#1}}}

\bibitem[{{Agol}(2011)}]{agol}
{Agol}, E. 2011, \apjl, 731, L31, \dodoi{10.1088/2041-8205/731/2/L31}

\bibitem[{{Alexander}(1973)}]{alex1973}
{Alexander}, M.~E. 1973, \apss, 23, 459, \dodoi{10.1007/BF00645172}

\bibitem[{{Anderson} {et~al.}(2016){Anderson}, {Storch}, \& {Lai}}]{ander2016}
{Anderson}, K.~R., {Storch}, N.~I., \& {Lai}, D. 2016, \mnras, 456, 3671,
  \dodoi{10.1093/mnras/stv2906}
nord
\bibitem[{{Bonsor} \& {Veras}(2015)}]{bonsor}
{Bonsor}, A., \& {Veras}, D. 2015, \mnras, 454, 53,
  \dodoi{10.1093/mnras/stv1913}

\bibitem[{{Cort{\'e}s} \& {Kipping}(2019)}]{CK2019}
{Cort{\'e}s}, J., \& {Kipping}, D. 2019, \mnras, 488, 1695,
  \dodoi{10.1093/mnras/stz1300}

\bibitem[{{Cummings} {et~al.}(2018){Cummings}, {Kalirai}, {Tremblay},
  {Ramirez-Ruiz}, \& {Choi}}]{cummings}
{Cummings}, J.~D., {Kalirai}, J.~S., {Tremblay}, P.~E., {Ramirez-Ruiz}, E., \&
  {Choi}, J. 2018, \apj, 866, 21, \dodoi{10.3847/1538-4357/aadfd6}

\bibitem[{{De Rosa} {et~al.}(2014){De Rosa}, {Patience}, {Wilson}, {Schneider},
  {Wiktorowicz}, {Vigan}, {Marois}, {Song}, {Macintosh}, {Graham}, {Doyon},
  {Bessell}, {Thomas}, \& {Lai}}]{derosa}
{De Rosa}, R.~J., {Patience}, J., {Wilson}, P.~A., {et~al.} 2014, \mnras, 437,
  1216, \dodoi{10.1093/mnras/stt1932}

\bibitem[{{Fabrycky} \& {Tremaine}(2007)}]{FT2007}
{Fabrycky}, D., \& {Tremaine}, S. 2007, \apj, 669, 1298, \dodoi{10.1086/521702}

\bibitem[{{Fernandes} {et~al.}(2019){Fernandes}, {Mulders}, {Pascucci},
  {Mordasini}, \& {Emsenhuber}}]{fernandes}
{Fernandes}, R.~B., {Mulders}, G.~D., {Pascucci}, I., {Mordasini}, C., \&
  {Emsenhuber}, A. 2019, \apj, 874, 81, \dodoi{10.3847/1538-4357/ab0300}

\bibitem[{{Ford} {et~al.}(2000){Ford}, {Kozinsky}, \& {Rasio}}]{ford2000}
{Ford}, E.~B., {Kozinsky}, B., \& {Rasio}, F.~A. 2000, \apj, 535, 385,
  \dodoi{10.1086/308815}

\bibitem[{{Fulton} {et~al.}(2014){Fulton}, {Tonry}, {Flewelling}, {Burgett},
  {Chambers}, {Hodapp}, {Huber}, {Kaiser}, {Wainscoat}, \& {Waters}}]{fulton}
{Fulton}, B.~J., {Tonry}, J.~L., {Flewelling}, H., {et~al.} 2014, \apj, 796,
  114, \dodoi{10.1088/0004-637X/796/2/114}

\bibitem[{{G{\"a}nsicke} {et~al.}(2019){G{\"a}nsicke}, {Schreiber}, {Toloza},
  {Gentile Fusillo}, {Koester}, \& {Manser}}]{gasincke}
{G{\"a}nsicke}, B.~T., {Schreiber}, M.~R., {Toloza}, O., {et~al.} 2019, \nat,
  576, 61, \dodoi{10.1038/s41586-019-1789-8}

\bibitem[{{Ghezzi} {et~al.}(2018){Ghezzi}, {Montet}, \& {Johnson}}]{occurrence}
{Ghezzi}, L., {Montet}, B.~T., \& {Johnson}, J.~A. 2018, \apj, 860, 109,
  \dodoi{10.3847/1538-4357/aac37c}

\bibitem[{{Guillochon} {et~al.}(2011){Guillochon}, {Ramirez-Ruiz}, \&
  {Lin}}]{guillo2011}
{Guillochon}, J., {Ramirez-Ruiz}, E., \& {Lin}, D. 2011, \apj, 732, 74,
  \dodoi{10.1088/0004-637X/732/2/74}

\bibitem[{{Hamer} \& {Schlaufman}(2020)}]{hamer2020}
{Hamer}, J.~H., \& {Schlaufman}, K.~C. 2020, \aj, 160, 138,
  \dodoi{10.3847/1538-3881/aba74f}

\bibitem[{{Hamers} \& {Lai}(2017)}]{HL2017}
{Hamers}, A.~S., \& {Lai}, D. 2017, \mnras, 470, 1657,
  \dodoi{10.1093/mnras/stx1319}

\bibitem[{{Hamers} \& {Portegies Zwart}(2016)}]{hamers2016}
{Hamers}, A.~S., \& {Portegies Zwart}, S.~F. 2016, \mnras, 462, L84,
  \dodoi{10.1093/mnrasl/slw134}

\bibitem[{{Howard} {et~al.}(2012){Howard}, {Marcy}, {Bryson}, {Jenkins},
  {Rowe}, {Batalha}, {Borucki}, {Koch}, {Dunham}, {Gautier}, {Van Cleve},
  {Cochran}, {Latham}, {Lissauer}, {Torres}, {Brown}, {Gilliland}, {Buchhave},
  {Caldwell}, {Christensen-Dalsgaard}, {Ciardi}, {Fressin}, {Haas}, {Howell},
  {Kjeldsen}, {Seager}, {Rogers}, {Sasselov}, {Steffen}, {Basri},
  {Charbonneau}, {Christiansen}, {Clarke}, {Dupree}, {Fabrycky}, {Fischer},
  {Ford}, {Fortney}, {Tarter}, {Girouard}, {Holman}, {Johnson}, {Klaus},
  {Machalek}, {Moorhead}, {Morehead}, {Ragozzine}, {Tenenbaum}, {Twicken},
  {Quinn}, {Isaacson}, {Shporer}, {Lucas}, {Walkowicz}, {Welsh}, {Boss},
  {Devore}, {Gould}, {Smith}, {Morris}, {Prsa}, {Morton}, {Still}, {Thompson},
  {Mullally}, {Endl}, \& {MacQueen}}]{howard2012}
{Howard}, A.~W., {Marcy}, G.~W., {Bryson}, S.~T., {et~al.} 2012, \apjs, 201,
  15, \dodoi{10.1088/0067-0049/201/2/15}

\bibitem[{{Hut}(1981)}]{hut1981}
{Hut}, P. 1981, \aap, 99, 126

\bibitem[{{Ito} \& {Ohtsuka}(2019)}]{ito2019}
{Ito}, T., \& {Ohtsuka}, K. 2019, Monographs on Environment, Earth and Planets,
  7, 1, \dodoi{10.5047/meep.2019.00701.0001}

\bibitem[{{Katz} {et~al.}(2011){Katz}, {Dong}, \& {Malhotra}}]{katz2011}
{Katz}, B., {Dong}, S., \& {Malhotra}, R. 2011, \prl, 107, 181101,
  \dodoi{10.1103/PhysRevLett.107.181101}

\bibitem[{{Kozai}(1962)}]{koz62}
{Kozai}, Y. 1962, \aj, 67, 591, \dodoi{10.1086/108790}

\bibitem[{{Kratter} \& {Perets}(2012)}]{KP12}
{Kratter}, K.~M., \& {Perets}, H.~B. 2012, \apj, 753, 91,
  \dodoi{10.1088/0004-637X/753/1/91}

\bibitem[{{Lai}(1997)}]{lai1997}
{Lai}, D. 1997, \apj, 490, 847, \dodoi{10.1086/304899}

\bibitem[{{Lidov}(1962)}]{lid62}
{Lidov}, M.~L. 1962, \planss, 9, 719, \dodoi{10.1016/0032-0633(62)90129-0}

\bibitem[{{Lithwick} \& {Naoz}(2011)}]{lithwick2011}
{Lithwick}, Y., \& {Naoz}, S. 2011, \apj, 742, 94,
  \dodoi{10.1088/0004-637X/742/2/94}

\bibitem[{{Liu} {et~al.}(2015){Liu}, {Mu{\~n}oz}, \& {Lai}}]{liu15}
{Liu}, B., {Mu{\~n}oz}, D.~J., \& {Lai}, D. 2015, \mnras, 447, 747,
  \dodoi{10.1093/mnras/stu2396}

\bibitem[{{Mardling}(1995)}]{mard1995}
{Mardling}, R.~A. 1995, \apj, 450, 732, \dodoi{10.1086/176179}

\bibitem[{{McCook} \& {Sion}(1999)}]{mccook1999}
{McCook}, G.~P., \& {Sion}, E.~M. 1999, \apjs, 121, 1, \dodoi{10.1086/313186}

\bibitem[{{Mu{\~n}oz} {et~al.}(2016){Mu{\~n}oz}, {Lai}, \& {Liu}}]{munoz2016}
{Mu{\~n}oz}, D.~J., {Lai}, D., \& {Liu}, B. 2016, \mnras, 460, 1086,
  \dodoi{10.1093/mnras/stw983}

\bibitem[{{Naoz}(2016)}]{naoz2016}
{Naoz}, S. 2016, \araa, 54, 441, \dodoi{10.1146/annurev-astro-081915-023315}

\bibitem[{{Naoz} {et~al.}(2012){Naoz}, {Farr}, \& {Rasio}}]{naoz2012}
{Naoz}, S., {Farr}, W.~M., \& {Rasio}, F.~A. 2012, \apjl, 754, L36,
  \dodoi{10.1088/2041-8205/754/2/L36}

\bibitem[{{Petrovich}(2015)}]{petro2015}
{Petrovich}, C. 2015, \apj, 799, 27, \dodoi{10.1088/0004-637X/799/1/27}

\bibitem[{{Petrovich} \& {Mu{\~n}oz}(2017)}]{petro2017}
{Petrovich}, C., \& {Mu{\~n}oz}, D.~J. 2017, \apj, 834, 116,
  \dodoi{10.3847/1538-4357/834/2/116}

\bibitem[{{Press} \& {Teukolsky}(1977)}]{press1977}
{Press}, W.~H., \& {Teukolsky}, S.~A. 1977, \apj, 213, 183,
  \dodoi{10.1086/155143}

\bibitem[{{Ronco} {et~al.}(2020){Ronco}, {Schreiber}, {Giuppone}, {Veras},
  {Cuadra}, \& {Guilera}}]{ronco}
{Ronco}, M.~P., {Schreiber}, M.~R., {Giuppone}, C.~A., {et~al.} 2020, \apjl,
  898, L23, \dodoi{10.3847/2041-8213/aba35f}

\bibitem[{{Shappee} \& {Thompson}(2013)}]{shappee2012}
{Shappee}, B.~J., \& {Thompson}, T.~A. 2013, \apj, 766, 64,
  \dodoi{10.1088/0004-637X/766/1/64}

\bibitem[{{Stephan} {et~al.}(2018){Stephan}, {Naoz}, \& {Gaudi}}]{stephan2018}
{Stephan}, A.~P., {Naoz}, S., \& {Gaudi}, B.~S. 2018, \aj, 156, 128,
  \dodoi{10.3847/1538-3881/aad6e5}

\bibitem[{{Stephan} {et~al.}(2017){Stephan}, {Naoz}, \&
  {Zuckerman}}]{stephan2017}
{Stephan}, A.~P., {Naoz}, S., \& {Zuckerman}, B. 2017, \apjl, 844, L16,
  \dodoi{10.3847/2041-8213/aa7cf3}

\bibitem[{{van Sluijs} \& {Van Eylen}(2018)}]{eylen}
{van Sluijs}, L., \& {Van Eylen}, V. 2018, \mnras, 474, 4603,
  \dodoi{10.1093/mnras/stx3068}

\bibitem[{{Vanderburg} {et~al.}(2020){Vanderburg}, {Rappaport}, {Xu},
  {Crossfield}, {Becker}, {Gary}, {Murgas}, {Blouin}, {Kaye}, {Palle}, {Melis},
  {Morris}, {Kreidberg}, {Gorjian}, {Morley}, {Mann}, {Parviainen}, {Pearce},
  {Newton}, {Carrillo}, {Zuckerman}, {Nelson}, {Zeimann}, {Brown},
  {Tronsgaard}, {Klein}, {Ricker}, {Vand erspek}, {Latham}, {Seager}, {Winn},
  {Jenkins}, {Adams}, {Benneke}, {Berardo}, {Buchhave}, {Caldwell},
  {Christiansen}, {Collins}, {Col{\'o}n}, {Daylan}, {Doty}, {Doyle},
  {Dragomir}, {Dressing}, {Dufour}, {Fukui}, {Glidden}, {Guerrero}, {Guo},
  {Heng}, {Henriksen}, {Huang}, {Kaltenegger}, {Kane}, {Lewis}, {Lissauer},
  {Morales}, {Narita}, {Pepper}, {Rose}, {Smith}, {Stassun}, \&
  {Yu}}]{vand2020}
{Vanderburg}, A., {Rappaport}, S.~A., {Xu}, S., {et~al.} 2020, arXiv e-prints,
  arXiv:2009.07282.
\newblock \doarXiv{2009.07282}

\bibitem[{{Veras}(2016)}]{veras_review}
{Veras}, D. 2016, Royal Society Open Science, 3, 150571,
  \dodoi{10.1098/rsos.150571}

\bibitem[{{Veras} \& {Fuller}(2020)}]{VF2020}
{Veras}, D., \& {Fuller}, J. 2020, \mnras, 492, 6059,
  \dodoi{10.1093/mnras/staa309}

\bibitem[{{Veras} {et~al.}(2013){Veras}, {Mustill}, {Bonsor}, \&
  {Wyatt}}]{veras2013}
{Veras}, D., {Mustill}, A.~J., {Bonsor}, A., \& {Wyatt}, M.~C. 2013, \mnras,
  431, 1686, \dodoi{10.1093/mnras/stt289}

\bibitem[{{Vick} \& {Lai}(2018)}]{vick2018}
{Vick}, M., \& {Lai}, D. 2018, \mnras, 476, 482, \dodoi{10.1093/mnras/sty225}

\bibitem[{{Vick} {et~al.}(2019){Vick}, {Lai}, \& {Anderson}}]{vick2019}
{Vick}, M., {Lai}, D., \& {Anderson}, K.~R. 2019, \mnras, 484, 5645,
  \dodoi{10.1093/mnras/stz354}

\bibitem[{{Villaver} {et~al.}(2014){Villaver}, {Livio}, {Mustill}, \&
  {Siess}}]{villaver}
{Villaver}, E., {Livio}, M., {Mustill}, A.~J., \& {Siess}, L. 2014, \apj, 794,
  3, \dodoi{10.1088/0004-637X/794/1/3}

\bibitem[{{von Zeipel}(1910)}]{zeipel1910}
{von Zeipel}, H. 1910, Astronomische Nachrichten, 183, 345,
  \dodoi{10.1002/asna.19091832202}

\bibitem[{{Wu}(2018)}]{wu2018}
{Wu}, Y. 2018, \aj, 155, 118, \dodoi{10.3847/1538-3881/aaa970}

\bibitem[{{Wu} \& {Murray}(2003)}]{wu2003}
{Wu}, Y., \& {Murray}, N. 2003, \apj, 589, 605, \dodoi{10.1086/374598}

\end{thebibliography}

\end{document}